\documentclass[ ]{ipart_v1}

\usepackage{t1enc}
\usepackage[latin1]{inputenc}
\usepackage[english]{babel}

\usepackage{amsthm}
\usepackage{yfonts}
\let\ltxcup\cup
\usepackage{bbm}
\usepackage{bm}
\usepackage{mathrsfs}
\usepackage{mathtools}

\newcommand{\comment}[1]{ \ding{118} \ding{118} \ding{118} #1  \ding{118} \ding{118} \ding{118}}
\newcommand{\bra}[1]{\langle #1|}
\newcommand{\ket}[1]{| #1\rangle}
\newcommand{\braket}[2]{\langle #1|#2\rangle}
\newcommand{\be}[0]{\begin{equation}}
\newcommand{\ee}[0]{\end{equation}}

\numberwithin{equation}{section}

\theoremstyle{plain}

\begin{document}

\title[Effects of hybridization and spin-orbit coupling]{Effects of hybridization and spin-orbit coupling  to induce odd frequency pairing in two-band superconductors}

\author[Moloud Tamadonpour and Heshmatollah Yavari]{Moloud Tamadonpour and Heshmatollah Yavari}

\begin{abstract}
The effects of spin independent hybridization potential and spin-orbit coupling on two-band superconductor with equal time s-wave interband pairing order parameter is investigated theoretically. To study symmetry classes in two-band superconductors the Gor'kov equations are solved analytically. By defining spin singlet and spin triplet s-wave order parameter due to two-band degree of freedom the symmetry classes of Cooper pair are studied. For spin singlet case it is shown that spin independent hybridization generates Cooper pair belongs to even-frequency spin singlet even-momentum even-band parity (ESEE) symmetry class for both intraband and interband pairing correlations. For spin triplet order parameter, intraband pairing correlation generates odd-frequency spin triplet even-momentum even-band parity (OTEE) symmetry class whereas, interband pairing correlation generates even-frequency spin triplet even-momentum odd-band parity (ETEO) class. For the spin singlet, spin-orbit coupling generates pairing correlation that belongs to odd-frequency spin singlet odd-momentum even-band parity (OSOE) symmetry class and even-frequency spin singlet even-momentum even-band parity (ESEE) for intraband and interband pairing correlation respectively. In the spin triplet case for itraband and interband correlation, spin-orbit coupling generates even-frequency spin triplet odd-momentum even-band parity (ETOE) and even-frequency spin triplet even-momentum odd-band parity (ETEO) respectively.
\end{abstract}

\maketitle

\section{Introduction and summary}

Symmetries of order parameter in superconductors affect their physical properties. The total wave function of a pair of fermions, in accordance with the Pauli principle, should be asymmetric under the permutation of orbital, spin and time (or equivalently Matsubara frequency) coordinates \cite{Sigrist:1991}. This leads to four classes allowed combinations for the symmetries of the wave function. This would imply that if the pairing is even in time, spin singlet pairs have even parity (ESE) and spin triplet pairs have odd parity (ETO). While if the pairing is odd in time, spin singlet pairs have odd parity (OSO) and spin triplet pairs have even parity (OTE). Black-Schaffer and Balatsky \cite{Black-Schafer and Balatsky:2013} have shown that the multiband superconducting order parameter has an extra symmetry classification that originates from the band degree of freedom, so called even-band-parity and odd band-parity. As a consequence, Cooper pairs can be classified into eight symmetry classes \cite{Asano:2015}.

Transport properties of multi-band superconductor are qualitatively different from those of the one-band superconductor. For instance, two-band system with the non-magnetic impurity violates Anderson theorem \cite{Moskalenko:1996}. As a result, lots of efforts have been devoted to understanding the properties of such systems both theoretically and experimentally. For these materials band symmetry plays important role. A main hypothesis of the model is the formation of the Cooper pairs inside one energy band and transition of this pair from one band to another which leads to intra and inter band electronic interactions. Multi-band model explained lots of strange physical properties of superconductive systems and were consistent with experimental data. Famous multiband superconductors are MgB2 \cite{Xi:2008,Nagamatsu:2001} and the iron-based superconductors	\cite{Shun-Li:2013,Tanaka:2010,Sprau:2013}. The nature of their two bands requires that the multiband approach be used to describe their properties. On the contrary for cuprates despite their multiband nature a single-band approach is more appropriate. 

From a general symmetry analysis of even and odd-frequency pairing states, it was shown that odd-frequency pairing always exists in the form of odd-interband (orbital) pairing if there is any even-frequency even-interband pairing present consistent with the general symmetry requirements \cite{Black-Schafer:2013}. The appearance of odd-frequency Cooper pairs in two-band superconductors by solving the Gor'kov equation was discussed analytically \cite{Asano and Golubov:2018}. They considered the equal-time s-wave pair potential and introduced two types of hybridization potentials between the two conduction bands. One is a spin-independent hybridization potential and the other is a spin-dependent hybridization potential derived from the spin-orbit interaction. 

The effect of random nonmagnetic impurities on the superconducting transition temperature in a two-band superconductor, by assuming the equal-time spin-singlet s-wave pair potential in each conduction band and the hybridization between the two bands as well as the band asymmetry was studied theoretically \cite{Asano and Golubov:2018,Asano:2018}. The effect of single-quasiparticle hybridization or scattering in a two-band superconductor by performing perturbation theory to infinite order in the hybridization term, in a multiband superconductor was investigated \cite{Komendov'a:2015}.

The superconducting state of multi-orbital spin-orbit coupled systems in the presence of an orbitally driven inversion asymmetry, by assuming that the interorbital attraction is the dominant pairing channel, was studied \cite{Fukaya:2018}. They have shown that in the absence of the inversion symmetry, superconducting states that avoid mixing of spin-triplet and spin-singlet configurations are allowed, and remarkably, spin-triplet states that are topologically nontrivial can be stabilized in a large portion of the phase diagram. The impact of strong spin-orbit coupling (SOC) on the properties of new class superconductors has attracted much attentions. It has been the subject of great theoretical and experimental interest \cite{Mineev:2012,Smidman:2017}. The formation of unexpected multi-component superconductors states allows for superconductors with magnetism and SOC. It was shown that for multi-orbital systems such as the Fe-pnictides SOC coupling, is much smaller than the orbit Hund's coupling \cite{Lee:2008,Yang:2009,Fazekas:1999}, In contrast for multiband systems such as Ir-based oxide materials it was found that the SOC interaction is comparable to the on-site Coulomb interaction \cite{Kuriyama:2010}. The combined effect of Hund's and SOC coupling on superconductivity in multi-orbital systems was investigated and it was shown that Hund's interaction leads to orbital-singlet spin-triplet superconductivity, where the Cooper pair wave function is antisymmetric under the exchange of two orbitals \cite{Christoph:2012}. Combined effect of the spin-orbit coupling and scattering on the nonmagnetic disorder on the formation of the spin resonance peak in iron-based superconductors was also studied \cite{Korshunov:2018}.

In this paper by using Gor'kov equation the effects of spin-orbit coupling and hybridization on the possibility of odd frequency pairing of a two-band superconductor with an equal time s-wave interband pairing order parameter are investigated theoretically.

\section{Formalism}

\subsection{Two-band model}

\label{SecModel}

The basic physics of multiband superconductors can be obtained by introducing a two-band model. We start with a normal two-band Hamiltonian as \cite{Asano:2018}
\be
\label{Eq1}
\check {\rm H}_N   = \int {dr\left[ {\begin{array}{*{20}c}
   {\psi ^\dag  _{1, \uparrow } (r),} & {\psi ^\dag  _{1, \downarrow } (r),} & {\psi ^\dag  _{2, \uparrow } (r),} & {\psi ^\dag  _{2, \downarrow } (r)}  \\
\end{array}} \right]} \check H _N (r)\left[ {\begin{array}{*{20}c}
   {\psi _{1, \uparrow } (r)}  \\
   {\psi _{1, \downarrow } (r)}  \\
   {\psi _{2, \uparrow } (r)}  \\
   {\psi _{2, \downarrow } (r)}  \\
\end{array}} \right],
\ee
where
\be
\label{Eq2}
\check H_N = \left( {\begin{array}{*{20}c}
   {\xi _{1k} \hat \sigma _0 } & {\left( {\upsilon e^{i\theta }  + V} \right)\hat \sigma _0 }  \\
   {\left( {\upsilon e^{ - i\theta }  + V^* } \right)\hat \sigma _0 } & {\xi _{2k} \hat \sigma _0 }  \\
\end{array}} \right).
\ee
Here $\psi _{\alpha ,\sigma } (r)$ is the annihilation ($\psi _{\alpha ,\sigma }^\dag  (r))$ creation) operator of an electron with spin ($\sigma  =  \uparrow , \downarrow $) at the $\alpha$ th conduction band,  $\xi _{\alpha k}  = \hbar ^2 k^2 /2m_e  - \mu _F $ is the dispersion energy of band $\alpha$, $m_e$ is  the mass of an electron, $\mu _F $ is the chemical potential. The spin independent hybridization potential is a complex number characterized by a phase $\theta$.  $\upsilon e^{i\theta } $ denotes the hybridization between the two bands, which is much smaller than the Fermi energy in the two conduction bands. In the absence of spin flip hybridization the spin-orbit coupling potential is $ V(k) = \eta \hat z.(\sigma  \times \vec k) = \eta (k_y \sigma _x  - k_x \sigma _y )$, where $ \eta $ is the parameter that describes the strength of the Rashba spin-orbit coupling and $\hat z$ is the unit vector perpendicular to the superconducting surface. This potential is odd-momentum-parity functions satisfying $V(k) =  - V( - k)$. 
Throughout this paper, Pauli matrices in spin, two-band, particle- hole spaces are respectively denoted by $ \hat \sigma _j $, $ \hat \rho _j $ and $ \hat \tau _j $ for $ j=1-3 $.
Superconducting order parameter in band $\alpha$ is:
\be
\label{Eq3}
\hat \Delta _{\alpha \alpha '} (k) = \left( {\begin{array}{*{20}c}
   {\Delta _{11} (k)} & {\Delta _{12} (k)}  \\
   {\Delta _{21} (k)} & {\Delta _{22} (k)}  \\
\end{array}} \right).
\ee
We focus only on interband superconducting order parameter $(\Delta _{11} (k) = \Delta _{22} (k) = 0)$.
The interband s-wave pair potential, is defined by  \cite{Asano:2018}
\be
\label{Eq4}
\Delta _{12,\sigma \sigma '} (r) = g\left\langle {\psi _{1,\sigma } (r)\psi _{2,\sigma '} (r)} \right\rangle .
\ee
here $g$ is interband attractive interaction between two electrons.
By assuming the spatially uniform order parameter the Fourier transformation of the pair potential becomes
\be
\label{Eq5}
\Delta _{12, \uparrow  \downarrow }  = \frac{g}{{V_{vol} }}\sum\limits_k {\left\langle {\psi _{1, \uparrow } (k)\psi _{2, \downarrow } ( - k)} \right\rangle } .
\ee
In the two-band model, for spin singlet the order parameter is symmetric (antisymmetric) under the permutation of band (spin) indices.
\be
\label{Eq6}
\Delta _{12, \uparrow  \downarrow }  = \Delta _{21; \uparrow  \downarrow }  =  - \Delta _{1,2; \downarrow  \uparrow } ,
\ee
But for spin triplet the order parameter is antisymmetric (symmetric) under the permutation of band (spin) indices.
\be
\label{Eq7}
\Delta _{12, \uparrow  \downarrow }  =  - \Delta _{21; \uparrow  \downarrow }  = \Delta _{1,2; \downarrow  \uparrow } .
\ee
For simplicity we omit the indices of $\Delta _{\alpha \alpha '} $.
The Hamiltonian describing superconductor in the Nambu space, can be written as \cite{Asano and Golubov:2018}
\be
\label{Eq8}
 \mathord{\buildrel{\lower3pt\hbox{$\scriptscriptstyle\smile$}} 
\over H} _{S(T)}  =\frac{1}{2}\sum\limits_k {\psi _{k,\sigma }^\dag  } \left( {\begin{array}{*{20}c}
   {\check H _N (k)} & {\check \Delta _{S(T)} }  \\
   {\check \Delta _{S(T)}^\dag  } & { - \check H  ^* _N ( - k)}  \\
\end{array}} \right)\psi _{k,\sigma } ,
\ee
where the spin-singlet and spin triplet pair potentials ($ \check \Delta _S $ and $\check \Delta _T $) are respectively given by
\be
\label{Eq9}
\check \Delta _S  = \Delta \hat \rho _1 i\hat \sigma _2 ,
\ee
\be
\label{Eq10}
\check \Delta _T  = \Delta i\hat \rho _2 \hat \sigma _1 .
\ee
For a two-band system, the Bogoliubov- de Gennes Hamiltonian can be described by $8 \times 8$ matrix reflecting spin, particle- hole and two band degrees of freedom. 
In particle-hole space $ N_1 $, by considering the spin of electron as $\uparrow $ and for hole as $\downarrow$,  while in particle-hole space $N_2$, we consider the spin of electron as $ \downarrow $ and for hole as $\uparrow$, we can describe the Hamiltonian $\check H_{S(T)}$ by a $4 \times 4$ matrix  \cite{Asano:2015,Asano:2018}
\be
\label{Eq11}
 \check H _0  =
\begin {pmatrix}
\xi _k & \upsilon e^{i\theta }  + V(k) & 0 & \Delta\\
\upsilon e^{ - i\theta }  + V^* (k) & \xi _k &  - s_{spin} \Delta & 0\\
 0 & - s_{spin} \Delta & - \xi _k  & - \upsilon e^{ - i\theta }  - V^* ( - k) \\
\Delta & 0 & - \upsilon e^{i\theta } - V( - k) &  - \xi _k \\
\end {pmatrix} 
\ee
here $ s_{spin}=-1 $ for spin singlet and $s_{spin}=1 $ for spin triplet.
To discuss the effects of hybridizations and spin-orbit interaction on the properties of superconductors, we calculate the Green's functions by solving the Gor'kov equation \cite{Gor'kov:1960}
\be
\label{Eq12}
\left( {i\omega _n  - \check H _0 } \right)\check G _0 (k,i\omega _n ) = \check 1 ,
\ee
\be
\label{Eq13}
\check G _0 (k,i\omega _n ) = \left( {\begin{array}{*{20}c}
   {\hat G_0 (k,i\omega _n )} & {\hat F_0 (k,i\omega _n )}  \\
   { - s_{spin} \hat F_0^\dag  ( - k,i\omega _n )} & { - \hat G_0^* ( - k,i\omega _n )}  \\
\end{array}} \right) .
\ee
where $ \omega _n  = \left( {2n + 1} \right)\pi k_B T $ is the Matsubara frequency ( $ k_{B} $ is the Boltzmann constant), and $ \check{1} $ is the identity matrix in $ spin \times band \times particle-hole $ space. $ \check {G_{0}} $ is a $ 4 \times 4 $ matrix where the diagonal components are normal Green's function and non-diagonal components are anomalous Green's function.

\subsection{Spin Singlet Pairing Order}

\label{SecSpinSinglet}

According to Equation \eqref{Eq11}, the Hamiltonian of a two-band superconductor with spin singlet configuration in the presence of spin-orbit coupling is 
\be
\label{Eq14}
 \check H _0  =
\begin {psmallmatrix}
\xi _k & \upsilon e^{i\theta }  + \eta (k_y  + ik_x ) & 0 & \Delta\\
\upsilon e^{ - i\theta }  + \eta (k_y  - ik_x ) & \xi _k & \Delta & 0\\
0 & \Delta & - \xi _k & - \upsilon e^{ - i\theta }  + \eta (k_y  - ik_x ) \\
\Delta & 0 & - \upsilon e^{i\theta }  + \eta (k_y  + ik_x ) &  - \xi _k\\
\end {psmallmatrix} .
\ee
By using Equation \eqref{Eq12} and \eqref{Eq13} , for spin singlet the solution of the normal Green's function within the first order of $ \Delta $ is calculated as
\be
\label{Eq15}
\begin{array}{l}
 \hat G_0 (k,i\omega _n ) = \frac{\Delta }{{Z_0 }}\{ [\left( {\xi  - i\omega _n } \right)(\nu ^2  + \eta ^2 k^2  - 2\nu \eta (k_x \sin \theta  + k_y \cos \theta ) + \left( {\xi  + i\omega _n } \right) \\ 
\qquad\qquad\:\:\:\: \times \left( { - \xi ^2  - \omega _n ^2 } \right)]\hat \rho _0  + [( - \nu \cos \theta  - \eta k_y )( - \left( {\xi  + i\omega _n } \right)^2  + \nu ^2  + \eta ^2 k^2  \\ 
\qquad\qquad\:\:\:\:- 2\nu \eta (k_x \sin \theta  + k_y \cos \theta ))]\hat \rho _1  + [(\nu \sin \theta  + \eta k_x )( - \left( {\xi  + i\omega _n } \right)^2    \\ 
\qquad\qquad\:\:\:\: + \nu ^2  + \eta ^2 k^2- 2\nu \eta (k_x \sin \theta  + k_y \cos \theta ))]\hat \rho _2 \}  \\ 
 \end{array}
\ee
here
\be
\label{Eq16}
\begin{array}{l}
 Z_0  = \xi ^4  + 2\xi ^2 \left( {\omega _n ^2  - \nu ^2 } \right) + \left( {\omega _n ^2  + \nu ^2 } \right)^2  - 8i\eta \nu \xi \omega _n (k_x \sin \theta  + k_y \cos \theta ) \\ 
\quad \: \: + 2\cos 2\theta \eta ^2 \nu ^2 (k_x ^2  - k_y ^2 ) - 4\sin 2\theta \eta ^2 \nu ^2 k_x k_y  + 2\eta ^2 k^2 \left( {\omega _n ^2  - \xi ^2 } \right) + \eta ^4 k^4 . \\ 
 \end{array} 
\ee
that $ k_{x}= k \cos \phi $ and $ k_{y} =k \sin \phi $, where $\phi$ is the angle between momentum and the $ x $ axis.
The matrix form of the normal Green's function ( Eq. \eqref{Eq15}) can be written as
\be
\label{Eq17}
\hat G_0 (k,i\omega _n ) = \left( {\begin{array}{*{20}c}
   {G_{11} (k,i\omega _n )} & {G_{12} (k,i\omega _n )}  \\
   {G_{21} (k,i\omega _n )} & {G_{22} (k,i\omega _n )}  \\
\end{array}} \right) .
\ee
where
\be
\label{Eq18}
\begin{array}{l}
 G_{11} (k,i\omega _n ) = \frac{\Delta }{{Z_0 }}[\left( {\xi  - i\omega _n } \right)(\nu ^2  + \eta ^2 (k_x ^2  + k_y ^2 ) - 2\nu \eta (k_x \sin \theta  + k_y \cos \theta ) \\ 
\qquad\qquad\quad\: - \left( {\xi  + i\omega _n } \right)(\xi ^2  + \omega _n ^2 )] \\ 
 \end{array}
\ee
\be
\label{Eq19}
\begin{array}{l}
 G_{12} (k,i\omega _n ) = \frac{\Delta }{{Z_0 }}[\{ - \nu e^{i\theta }  - \eta (ik_x  + k_y )\} \{ - \left( {\xi  + i\omega _n } \right)^2  + \nu ^2  + \eta ^2 k^2  \\ 
\qquad\qquad\quad\:- 2\nu \eta (k_x \sin \theta  + k_y \cos \theta )\}] \\ 
 \end{array}
\ee
\be
\label{Eq20}
\begin{array}{l}
 G_{21} (k,i\omega _n ) = \frac{\Delta }{{Z_0 }}[\{ - \nu e^{ - i\theta }  - \eta ( - ik_x  + k_y )\} \{  - \left( {\xi  + i\omega _n } \right)^2  + \nu ^2  + \eta ^2 k^2  \\ 
\qquad\qquad\quad\: - 2\nu \eta (k_x \sin \theta  + k_y \cos \theta )\} ] \\ 
 \end{array}
\ee
\be
\label{Eq21}
\begin{array}{l}
 G_{22} (k,i\omega _n ) = \frac{\Delta }{{Z_0 }}[\left( {\xi  - i\omega _n } \right)(\nu e^{ - i\theta }  - \eta ( - ik_x  + k_y ))(\nu e^{i\theta }  - \eta (ik_x  + k_y ) \\ 
\qquad\qquad\quad\:- \left( {\xi  + i\omega _n } \right)\left( {\xi ^2  + \omega _n ^2 } \right)] \\ 
 \end{array}
\ee
By using Equation \eqref{Eq12}  and \eqref{Eq13}, the anomalous Green's function can be obtained as
\be
\label{Eq22}
\begin{array}{l}
 \hat F_0 (k,i\omega _n ) = \frac{\Delta }{{Z_0 }}[\left( {2\nu \xi \cos \theta  + 2\eta \omega _n ik_y } \right)\hat \rho _0  + \left( { - (\nu ^2  + \xi ^2  + \omega _n ^2 ) + \eta ^2 k^2 } \right)\hat \rho _1  \\ 
\qquad\qquad\quad\:+ \left( {2\nu \eta \left( {k_x co{\mathop{\rm s}\nolimits} \theta  - k_y \sin \theta } \right)} \right)\hat \rho _2  + \left( {2\nu \xi i\sin \theta  - 2\eta \omega _n k_x } \right)\hat \rho _3 ] . \\ 
 \end{array}
\ee
In particle- hole space $N_{1}$ , the matrix form of the anomalous Green's function (Eq. \eqref{Eq22}) is 
\be
\label{Eq23}
\hat F_0 ^{N_1 } (k,i\omega _n ) = \left( {\begin{array}{*{20}c}
   {F_{11, \uparrow  \downarrow } (k,i\omega _n )} & {F_{12, \uparrow  \downarrow } (k,i\omega _n )}  \\
   {F_{21, \uparrow  \downarrow } (k,i\omega _n )} & {F_{22, \uparrow  \downarrow } (k,i\omega _n )}  \\
\end{array}} \right) ,
\ee
where
\be
\label{Eq24}
F_{11, \uparrow  \downarrow } (k,i\omega _n ) = \frac{\Delta }{{Z_0 }}[2\nu \xi e^{i\theta }  - 2\eta \omega _n k_x  + 2\eta \omega _n ik_y ] ,
\ee
\be
\label{Eq25}
F_{12, \uparrow  \downarrow } (k,i\omega _n ) = \frac{\Delta }{{Z_0 }}[ - (\nu ^2  + \xi ^2  + \omega _n ^2 ) + \eta ^2 k^2  - 2i\nu \eta (k_x co{\mathop{\rm s}\nolimits} \theta  - k_y \sin \theta )] ,
\ee
\be
\label{Eq26}
F_{21, \uparrow  \downarrow } (k,i\omega _n ) = \frac{\Delta }{{Z_0 }}[ - (\nu ^2  + \xi ^2  + \omega _n ^2 ) + \eta ^2 k^2  + 2i\nu \eta (k_x co{\mathop{\rm s}\nolimits} \theta  - k_y \sin \theta )] ,
\ee
\be
\label{Eq27}
F_{22, \uparrow  \downarrow } (k,i\omega _n ) = \frac{\Delta }{{Z_0 }}[2\nu \xi e^{i\theta }  - 2\eta \omega _n k_x  + 2\eta \omega _n ik_y ] .
\ee
In particle- hole space $N_{2}$, the matrix form of the anomalous Green's function is
\be
\label{Eq28}
\hat F_0 ^{N_2 } (k,i\omega _n ) = \left( {\begin{array}{*{20}c}
   {F_{11, \downarrow  \uparrow } (k,i\omega _n )} & {F_{12, \downarrow  \uparrow } (k,i\omega _n )}  \\
   {F_{21, \downarrow  \uparrow } (k,i\omega _n )} & {F_{22, \downarrow  \uparrow } (k,i\omega _n )}  \\
\end{array}} \right) =  - \hat F_0 ^{N_1 } (k,i\omega _n ) .
\ee

In the absence of spin-orbit coupling ($\eta$) the anomalous Green's function (Eq.\eqref{Eq22}) becomes
\be
\label{Eq29}
\hat F_0 (k,i\omega _n ) = \frac{\Delta }{{Z_0 }}[2\nu \xi \cos \theta \hat \rho _0  - (\nu ^2  + \xi ^2  + \omega _n ^2 )\hat \rho _1  + 2\nu \xi i\sin \theta \hat \rho _3 ] ,
\ee
here
\be
\label{Eq30}
Z_0  = \xi ^4  + 2\xi ^2 \left( {\omega _n ^2  - \nu ^2 } \right) + \left( {\omega _n ^2  + \nu ^2 } \right)^2 .
\ee
The matrix form of the anomalous Green's function in Equation \eqref{Eq29} in particle-hole spaces $N_{1}$ and $N_{2}$, are 
\be
\label{Eq31}
\begin{array}{l}
 \hat F_0 ^{N_1 } (k,i\omega _n ) = \left( {\begin{array}{*{20}c}
   {F_{11, \uparrow  \downarrow } (k,i\omega _n )} & {F_{12, \uparrow  \downarrow } (k,i\omega _n )}  \\
   {F_{21, \uparrow  \downarrow } (k,i\omega _n )} & {F_{22, \uparrow  \downarrow } (k,i\omega _n )}  \\
\end{array}} \right) \\ 
\qquad\qquad\quad\: = \frac{\Delta }{{Z_0 }}\left( {\begin{array}{*{20}c}
   {2\nu \xi \cos \theta  + 2i\nu \xi \sin \theta } & { - (\nu ^2  + \xi ^2  + \omega _n ^2 )}  \\
   { - (\nu ^2  + \xi ^2  + \omega _n ^2 )} & {2\nu \xi \cos \theta  - 2i\nu \xi \sin \theta }  \\
\end{array}} \right) \\ 
\qquad\qquad\quad\:= \frac{\Delta }{{Z_0 }}\left( {\begin{array}{*{20}c}
   {2\xi \nu e^{i\theta } } & { - (\nu ^2  + \xi ^2  + \omega _n ^2 )}  \\
   { - (\nu ^2  + \xi ^2  + \omega _n ^2 )} & {2\xi \nu e^{ - i\theta } }  \\
\end{array}} \right) , \\ 
 \end{array}
\ee
and
\be
\label{Eq32}
\begin{array}{l}
 \hat F_0 ^{N_2 } (k,i\omega _n ) = \left( {\begin{array}{*{20}c}
   {F_{11, \downarrow  \uparrow } (k,i\omega _n )} & {F_{12, \downarrow  \uparrow } (k,i\omega _n )}  \\
   {F_{21, \downarrow  \uparrow } (k,i\omega _n )} & {F_{22, \downarrow  \uparrow } (k,i\omega _n )}  \\
\end{array}} \right) =  - \hat F_0 ^{N_1 } (k,i\omega _n ) \\ 
\qquad\qquad\quad\: = \frac{\Delta }{{Z_0 }}\left( {\begin{array}{*{20}c}
   { - 2\nu \xi \cos \theta  - 2i\nu \xi \sin \theta } & {(\nu ^2  + \xi ^2  + \omega _n ^2 )}  \\
   {(\nu ^2  + \xi ^2  + \omega _n ^2 )} & { - 2\nu \xi \cos \theta  + 2i\nu \xi \sin \theta }  \\
\end{array}} \right) \\ 
\qquad\qquad\quad\: = \frac{\Delta }{{Z_0 }}\left( {\begin{array}{*{20}c}
   { - 2\xi \nu e^{i\theta } } & {(\nu ^2  + \xi ^2  + \omega _n ^2 )}  \\
   {(\nu ^2  + \xi ^2  + \omega _n ^2 )} & { - 2\xi \nu e^{ - i\theta } }  \\
\end{array}} \right) . \\ 
 \end{array}
\ee
The intraband pairing correlations become
\be
\label{Eq33}
F_{11, \uparrow  \downarrow } (k,i\omega _n ) - F_{11, \downarrow  \uparrow } (k,i\omega _n ) = \frac{{4\Delta }}{{Z_0 }}\xi \nu e^{i\theta } ,
\ee
\be
\label{Eq34}
F_{22, \uparrow  \downarrow } (k,i\omega _n ) - F_{22, \downarrow  \uparrow } (k,i\omega _n ) = \frac{{4\Delta }}{{Z_0 }}\xi \nu e^{ - i\theta } .
\ee
Hybridization generates $\rho_{0}$ and $\rho_{3}$ components which belongs to even frequency symmetry class. It means that in the presence of interband coupling, hybridization generates even frequency intra- sublattice pairing in the system. These components belong to even-frequency spin-singlet even-momentum even-band parity (ESEE) symmetry class. This result is in agreement with the equation (20) presented in Ref \cite{Asano:2018}. Equation \eqref{Eq33} and \eqref{Eq34} are in agreement with the Equation (62) and (63) reported in Ref \cite{Asano:2015} in the first order of $\Delta$ ($\lvert\Delta\rvert ^2=0$) and equal energy bands ($\xi _ -   = 0$) and both belong to the (ESEE) symmetry class.
The band symmetry generates interband pairing correlation:
\be
\label{Eq35}
\begin{array}{l}
 [F_{12, \uparrow  \downarrow } (k,i\omega _n ) - F_{12, \downarrow  \uparrow } (k,i\omega _n )] + [F_{21, \uparrow  \downarrow } (k,i\omega _n ) - F_{21, \downarrow  \uparrow } (k,i\omega _n )] \\ 
  = \frac{{ - 4\Delta }}{{Z_0 }}(\nu ^2  + \xi ^2  + \omega _n ^2 ) . \\ 
 \end{array}
\ee
which belongs to (ESEE). This result is in agreement with the Equation (65) presented in Ref \cite{Asano:2015}.
\be
\label{Eq36}
\begin{array}{l}
 [F_{12, \uparrow  \downarrow } (k,i\omega _n ) - F_{12, \downarrow  \uparrow } (k,i\omega _n )] - [F_{21, \uparrow  \downarrow } (k,i\omega _n ) - F_{21, \downarrow  \uparrow } (k,i\omega _n )] \\ 
  = \frac{{2\Delta }}{{Z_3 }}i\omega _n \xi _ -  .\\ 
 \end{array}
\ee
which belongs to the symmetry (OSEO) class.
We considered a two-band superconductor with an equal dispersion energy in each band $(\xi_{+}=\xi_{-})$. In this case the interband pairing correlation due to band asymmetry is
 \be
\label{Eq37}
[F_{12, \uparrow  \downarrow } (k,i\omega _n ) - F_{12, \downarrow  \uparrow } (k,i\omega _n )] - [F_{21, \uparrow  \downarrow } (k,i\omega _n ) - F_{21, \downarrow  \uparrow } (k,i\omega _n )] = 0 .
\ee

In the absence of hybridization within the second order of the spin-orbit coupling constant ($ \eta $), we obtain
\be
\label{Eq38}
\hat G_0 (k,i\omega _n ) = \frac{\Delta }{{Z_0 }}[(\eta ^2 k^2  - \left( {\xi  + i\omega _n } \right)^2 )\left( {\xi  - i\omega _n } \right)\hat \rho _0  + \eta \left( {k_y  + ik_x } \right)\left( {\xi  + i\omega _n } \right)^2 \hat \rho _1 ],
\ee
where
\be
\label{Eq39}
Z_0  = (\xi ^2  + \omega _n ^2 )^2  + 2\eta ^2 k^2 \left( {\omega _n ^2  - \xi ^2 } \right).
\ee
Equation \eqref{Eq22} can be rewritten as
\be
\label{Eq40}
\hat F_0 (k,i\omega ) = \frac{\Delta }{{Z_0 }}[2i\eta \omega _n k_y \hat \rho _0  - 2\eta \omega _n k_x \hat \rho _3  + \left( { - \xi ^2  - \omega _n ^2  + \eta ^2 k^2 } \right)\hat \rho _1 ].
\ee
In particle- hole space $ N_{1} $ and $ N_{2} $ ,the matrix form of the anomalous Green's function (Eq. \eqref{Eq40}) is 
\be
\label{Eq41}
\begin{array}{l}
 \hat F_0 ^{N_1 } (k,i\omega _n ) = \left( {\begin{array}{*{20}c}
   {F_{11, \uparrow  \downarrow } (k,i\omega _n )} & {F_{12, \uparrow  \downarrow } (k,i\omega _n )}  \\
   {F_{21, \uparrow  \downarrow } (k,i\omega _n )} & {F_{22, \uparrow  \downarrow } (k,i\omega _n )}  \\
\end{array}} \right) \\ 
\qquad\qquad\quad\:\: = \frac{\Delta }{{Z_0 }}\left( {\begin{array}{*{20}c}
   {2i\eta \omega _n k_y  - 2\eta \omega _n k_x } & { - \xi ^2  - \omega _n ^2  + \eta ^2 k^2 }  \\
   { - \xi ^2  - \omega _n ^2  + \eta ^2 k^2 } & {2i\eta \omega _n k_y  + 2\eta \omega _n k_x }  \\
\end{array}} \right) ,\\ 
 \end{array}
\ee
\be
\label{Eq42}
\begin{array}{l}
 \hat F_0 ^{N_2 } (k,i\omega _n ) = \left( {\begin{array}{*{20}c}
   {F_{11, \downarrow  \uparrow } (k,i\omega _n )} & {F_{12, \downarrow  \uparrow } (k,i\omega _n )}  \\
   {F_{21, \downarrow  \uparrow } (k,i\omega _n )} & {F_{22, \downarrow  \uparrow } (k,i\omega _n )}  \\
\end{array}} \right) =  - \hat F_0 ^{N_1 } (k,i\omega _n ) \\ 
\qquad\qquad\quad\: = \frac{\Delta }{{Z_0 }}\left( {\begin{array}{*{20}c}
   { - 2i\eta \omega _n k_y  + 2\eta \omega _n k_x } & {\xi ^2  + \omega _n ^2  - \eta ^2 k^2 }  \\
   {\xi ^2  + \omega _n ^2  - \eta ^2 k^2 } & { - 2i\eta \omega _n k_y  - 2\eta \omega _n k_x }  \\
\end{array}} \right) .\\ 
 \end{array}
\ee
The intraband pairing correlations are 
\be
\label{Eq43}
F_{11, \uparrow  \downarrow } (k,i\omega _n ) - F_{11, \downarrow  \uparrow } (k,i\omega _n ) = \frac{{4\Delta }}{{Z_0 }}i\omega _n \eta (k_y  + ik_x ),
\ee
\be
\label{Eq44}
F_{22, \uparrow  \downarrow } (k,i\omega _n ) - F_{22, \downarrow  \uparrow } (k,i\omega _n ) = \frac{{4\Delta }}{{Z_0 }}i\omega _n \eta (k_y  - ik_x ).
\ee
Spin-orbit coupling generates $\rho_{0}$ and $\rho_{3}$ components which belong to odd frequency symmetry class. It means that in the presence of inter-band coupling, spin-orbit coupling generates odd-frequency intrasublattice pairing in the system. These components belong to odd-frequency spin-singlet odd-momentum even-band parity (OSOE) symmetry class. 
In Ref \cite{Asano:2015} the intraband pairing correlation is written as
\be
\label{Eq45}
F_{11, \uparrow  \downarrow } (k,i\omega _n ) + F_{11, \downarrow  \uparrow } (k,i\omega _n ) = \frac{\Delta }{{Z_3 }}(\xi _ +   - \xi _ -  )V_3 ,
\ee
\be
\label{Eq46}
F_{22, \uparrow  \downarrow } (k,i\omega _n ) + F_{22, \downarrow  \uparrow } (k,i\omega _n ) = \frac{\Delta }{{Z_3 }}(\xi _ +   + \xi _ -  )V_3 .
\ee
The hybridization generates pairing correlations that belong to the (ETOE) class.
The band asymmetry generates interband pairing correlation as 
\be
\label{Eq47}
[F_{12, \uparrow  \downarrow } (k,i\omega _n ) - F_{12, \downarrow  \uparrow } (k,i\omega _n )] - [F_{21, \uparrow  \downarrow } (k,i\omega _n ) - F_{21, \downarrow  \uparrow } (k,i\omega _n )] = 0 .
\ee
In Ref \cite{Asano:2015}  for spin orbit hybridization the band asymmetry generates the interband pair correlation as 
\be
\label{Eq48}
[F_{12, \uparrow  \downarrow } (k,i\omega ) - F_{12, \downarrow  \uparrow } (k,i\omega )] - [F_{21, \uparrow  \downarrow } (k,i\omega ) - F_{21, \downarrow  \uparrow } (k,i\omega )] = \frac{{2\Delta }}{{Z_3 }}i\omega _n \xi _ -  .
\ee
which belongs to the odd-frequency spin-singlet even-momentum odd-band parity symmetry (OSEO).
The interband pairing correlation due to band symmetry is 
\be
\label{Eq49}
\begin{array}{l}
 [F_{12, \uparrow  \downarrow } (k,i\omega _n ) - F_{12, \downarrow  \uparrow } (k,i\omega _n )] + [F_{21, \uparrow  \downarrow } (k,i\omega _n ) - F_{21, \downarrow  \uparrow } (k,i\omega _n )] \\ 
  = \frac{{ - 4\Delta }}{{Z_0 }}(\xi ^2  + \omega _n ^2  - \eta ^2 k^2 ) .\\ 
 \end{array}
\ee
 This component belongs to even-frequency spin-singlet even-momentum even-band parity (ESEE) symmetry class. 
For spin singlet, hybridization potential generates ESEE symmetry class due to both intra and interband correlation, whereas the spin dependent hybridization potential generates this class only for interband pairing correlation due to band symmetry.
In this case the odd frequency pairing arises only due to intraband pairing correlations for spin dependent hybridization potential.

\subsection{Spin Triplet Pairing Order}

\label{SecSpinTriplet}

By considering Equation \eqref {Eq11}, the Hamiltonian of a two-band superconductor with spin triplet configuration in the presence of spin-orbit coupling is 
\be
\label{Eq50}
 \check H _0  =
\begin {psmallmatrix}
\xi _k & \upsilon e^{i\theta }  + \eta (k_y  + ik_x ) & 0 & \Delta\\
\upsilon e^{ - i\theta }  + \eta (k_y  - ik_x ) & \xi _k & -\Delta & 0\\
0 & -\Delta & - \xi _k & - \upsilon e^{ - i\theta }  + \eta (k_y  - ik_x ) \\
\Delta & 0 & - \upsilon e^{i\theta }  + \eta (k_y  + ik_x ) &  - \xi _k\\
\end {psmallmatrix} .
\ee
The solution of the anomalous Green's function within the first order of $ \Delta $ is calculated as
\be
\label{Eq51}
\begin{array}{l}
 \hat F_0 (k,i\omega _n ) = \frac{\Delta }{{Z_0 }}[( - 2i\eta \xi k_x  + 2\nu \omega _n \sin \theta )\hat \rho _0  + \left( {2i\nu \eta \left( {k_x co{\mathop{\rm s}\nolimits} \theta  - k_y \sin \theta } \right)} \right)\hat \rho _1  \\ 
\qquad\qquad\:\:\:+ (i(\nu ^2  - \xi ^2  - \omega _n ^2 ) - i\eta ^2 (k_x ^2  + k_y ^2 ))\hat \rho _2  + ( - 2\eta \xi k_y  - 2i\nu \omega _n \cos \theta )\hat \rho _3 ] . \\ 
 \end{array} 
\ee
The matrix form of the anomalous Green's function (Eq. \eqref {Eq51} ) can be written as
\be
\label{Eq52}
\hat F_{11, \uparrow  \downarrow } (k,i\omega _n ) = \frac{\Delta }{{Z_0 }}\left( { - 2i\nu \omega _n e^{i\theta }  - 2i\eta \xi k_x  - 2\eta \xi k_y } \right) ,
\ee
\be
\label{Eq53}
\hat F_{12, \uparrow  \downarrow } (k,i\omega _n ) = \frac{\Delta }{{Z_0 }}[(\nu ^2  - \xi ^2  - \omega _n ^2 ) - \eta ^2 k^2  + 2i\nu \eta \left( {k_x co{\mathop{\rm s}\nolimits} \theta  - k_y \sin \theta } \right)] ,
\ee
\be
\label{Eq54}
\hat F_{21, \uparrow  \downarrow } (k,i\omega _n ) = \frac{\Delta }{{Z_0 }}[( - \nu ^2  + \xi ^2  + \omega _n ^2 ) + \eta ^2 k^2  + 2i\nu \eta \left( {k_x co{\mathop{\rm s}\nolimits} \theta  - k_y \sin \theta } \right)] ,
\ee
\be
\label{Eq55}
\hat F_{22, \uparrow  \downarrow } (k,i\omega _n ) = \frac{\Delta }{{Z_0 }}\left( {2i\nu \omega _n e^{ - i\theta }  - 2i\eta \xi k_x  + 2\eta \xi k_y } \right) .
\ee

In the absence of spin-orbit coupling ($ \eta=0 $ ) the anomalous Green's function Equation \eqref{Eq51} becomes
\be
\label{Eq56}
\hat F_0 (k,i\omega _n ) = \frac{\Delta }{{Z_0 }}[2\nu \omega _n \sin \theta \hat \rho _0  + i(\nu ^2  - \xi ^2  - \omega _n ^2 )\hat \rho _2  - 2i\nu \omega _n \cos \theta \hat \rho _3 ] .
\ee
here
\be
\label{Eq57}
Z_0  = \xi ^4  + 2\xi ^2 \left( {\omega _n ^2  - \nu ^2 } \right) + \left( {\omega _n ^2  + \nu ^2 } \right)^2 .
\ee
In particle-hole space $ N_1 $ and $N_2 $ , the matrix form of the anomalous Green's function (Eq. \eqref {Eq56}) is 
\be
\label{Eq58}
\begin{array}{l}
 \hat F_0 ^{N_1 } (k,i\omega _n ) = \left( {\begin{array}{*{20}c}
   {F_{11, \uparrow  \downarrow } (k,i\omega _n )} & {F_{12, \uparrow  \downarrow } (k,i\omega _n )}  \\
   {F_{21, \uparrow  \downarrow } (k,i\omega _n )} & {F_{22, \uparrow  \downarrow } (k,i\omega _n )}  \\
\end{array}} \right) \\ 
\qquad\qquad\quad\:\:= \frac{\Delta }{{Z_0 }}\left( {\begin{array}{*{20}c}
   {2\nu \omega _n (\sin \theta  - i\cos \theta )} & {(\nu ^2  - \xi ^2  - \omega _n ^2 )}  \\
   { - (\nu ^2  - \xi ^2  - \omega _n ^2 )} & {2\nu \omega _n (\sin \theta  + i\cos \theta )}  \\
\end{array}} \right) , \\ 
 \end{array}
\ee
\be
\label{Eq59}
\begin{array}{l}
 \hat F_0 ^{N_2 } (k,i\omega _n ) = \left( {\begin{array}{*{20}c}
   {F_{11, \downarrow  \uparrow } (k,i\omega _n )} & {F_{12, \downarrow  \uparrow } (k,i\omega _n )}  \\
   {F_{21, \downarrow  \uparrow } (k,i\omega _n )} & {F_{22, \downarrow  \uparrow } (k,i\omega _n )}  \\
\end{array}} \right) = \hat F_0 ^{N_1 } (k,i\omega _n ) \\ 
\qquad\qquad\quad\:\:= \frac{\Delta }{{Z_0 }}\left( {\begin{array}{*{20}c}
   {2\nu \omega _n (\sin \theta  - ico{\mathop{\rm s}\nolimits} \theta )} & {(\nu ^2  - \xi ^2  - \omega _n ^2 )}  \\
   { - (\nu ^2  - \xi ^2  - \omega _n ^2 )} & {2\nu \omega _n (\sin \theta  + ico{\mathop{\rm s}\nolimits} \theta )} \\
\end{array}} \right) . \\ 
 \end{array}
\ee
The intraband pairing correlations becomes
\be
\label{Eq60}
F_{11, \uparrow  \downarrow } (k,i\omega _n ) + F_{11, \downarrow  \uparrow } (k,i\omega _n ) = \frac{{ - 4\Delta }}{{Z_0 }}i\omega _n \nu e^{ - i\theta } ,
\ee
\be
\label{Eq61}
F_{22, \uparrow  \downarrow } (k,i\omega _n ) + F_{22, \downarrow  \uparrow } (k,i\omega _n ) = \frac{{4\Delta }}{{Z_0 }}i\omega _n \nu e^{i\theta } .
\ee
Hybridization generates $ \rho_0 $ and $ \rho_3 $ which belongs to odd frequency symmetry class. These components belong to odd-frequency spin-triplet even-momentum even-band parity (OTEE) symmetry class. This result is in agreement with the Equation (24) presents in Ref \cite{Asano:2018} In the first order of $ \Delta $ ($\lvert\Delta\rvert ^2=0$ ) and equal energy bands ($\xi _ -   = 0$ ) Equation \eqref{Eq60} and \eqref{Eq61} are coincide with the Equation (83) and (84) presented in Ref \cite{Asano:2015} and both belong to the (OTEE) symmetry class.
The band symmetry generates interband pairing correlation as 
\be
\label{Eq62}
\begin{array}{l}
 [F_{12, \uparrow  \downarrow } (k,i\omega _n ) + F_{12, \downarrow  \uparrow } (k,i\omega _n )] - [F_{21, \uparrow  \downarrow } (k,i\omega _n ) + F_{21, \downarrow  \uparrow } (k,i\omega _n )] \\ 
  = \frac{{4\Delta }}{{Z_0 }}(\nu ^2  - \xi ^2  - \omega _n ^2 ) . \\ 
 \end{array}
\ee
which belongs to even-frequency spin triplet even-momentum odd-band parity (ETEO) symmetry class.
In Ref \cite{Asano:2015} the interband pairing correlation due to band asymmetry is 
\be
\label{Eq63}
\begin{array}{l}
 [F_{12, \uparrow  \downarrow } (k,i\omega _n ) + F_{12, \downarrow  \uparrow } (k,i\omega _n )] + [F_{21, \uparrow  \downarrow } (k,i\omega _n ) + F_{21, \downarrow  \uparrow } (k,i\omega _n )] \\ 
  = \frac{{2\Delta }}{{Z_5 }}i\omega _n \xi _ -   .\\ 
 \end{array}
\ee
Thus the band hybridization generates pairing correlations that belong to the odd-frequency spin triplet even-momentum even-band parity (OTEE) class.
Since we considered a two-band superconductor with an equal energy bands, the interband pairing correlation due to band asymmetry is
\be
\label{Eq64}
[F_{12, \uparrow  \downarrow } (k,i\omega _n ) + F_{12, \downarrow  \uparrow } (k,i\omega _n )] + [F_{21, \uparrow  \downarrow } (k,i\omega _n ) + F_{21, \downarrow  \uparrow } (k,i\omega _n )] = 0 .
\ee

In the absence of hybridization, we obtain
\be
\label{Eq65}
\hat F_0 (k,i\omega _n ) = \frac{\Delta }{{Z_0 }}[ - 2i\eta \xi k_x \hat \rho _0  - 2\eta \xi k_y \hat \rho _3  - i\left( {\xi ^2  + \omega _n ^2  + \eta ^2 k^2 } \right)\hat \rho _2 ] .
\ee
The matrix form of the anomalous Green's function in Equation \eqref{Eq65} in particle- hole spaces $ N_1 $ and $ N_2 $ , are
\be
\label{Eq66}
\begin{array}{l}
 \hat F_0 ^{N_1 } (k,i\omega _n ) = \left( {\begin{array}{*{20}c}
   {F_{11, \uparrow  \downarrow } (k,i\omega _n )} & {F_{12, \uparrow  \downarrow } (k,i\omega _n )}  \\
   {F_{21, \uparrow  \downarrow } (k,i\omega _n )} & {F_{22, \uparrow  \downarrow } (k,i\omega _n )}  \\
\end{array}} \right) \\ 
\qquad\qquad\quad\:\:= \frac{\Delta }{{Z_0 }}\left( {\begin{array}{*{20}c}
   { - 2i\eta \xi k_x  - 2\eta \xi k_y } & { - (\xi ^2  + \omega _n ^2  + \eta ^2 k^2 )}  \\
   {(\xi ^2  + \omega _n ^2  + \eta ^2 k^2 )} & { - 2i\eta \xi k_x  + 2\eta \xi k_y }  \\
\end{array}} \right) ,\\ 
 \end{array}
\ee\be
\label{Eq67}
\begin{array}{l}
 \hat F_0 ^{N_2 } (k,i\omega _n ) = \left( {\begin{array}{*{20}c}
   {F_{11, \downarrow  \uparrow } (k,i\omega _n )} & {F_{12, \downarrow  \uparrow } (k,i\omega _n )}  \\
   {F_{21, \downarrow  \uparrow } (k,i\omega _n )} & {F_{22, \downarrow  \uparrow } (k,i\omega _n )}  \\
\end{array}} \right) = \hat F_0 ^{N_1 } (k,i\omega _n ) \\ 
 \qquad\qquad\quad\:\:= \frac{\Delta }{{Z_0 }}\left( {\begin{array}{*{20}c}
   { - 2i\eta \xi k_x  - 2\eta \xi k_y } & { - (\xi ^2  + \omega _n ^2  + \eta ^2 k^2 )}  \\
   {(\xi ^2  + \omega _n ^2  + \eta ^2 k^2 )} & { - 2i\eta \xi k_x  + 2\eta \xi k_y }  \\
\end{array}} \right) .\\ 
 \end{array}
\ee
The intraband pairing correlations are 
\be
\label{Eq68}
F_{11, \uparrow  \downarrow } (k,i\omega _n ) + F_{11, \downarrow  \uparrow } (k,i\omega _n ) = \frac{{ - 4\Delta }}{{Z_0 }}\xi \eta (k_y  + ik_x ) ,
\ee
\be
\label{Eq69}
F_{22, \uparrow  \downarrow } (k,i\omega _n ) + F_{22, \downarrow  \uparrow } (k,i\omega _n ) = \frac{{4\Delta }}{{Z_0 }}\xi (k_y  - ik_x ) .
\ee
Spin-orbit coupling generates $ \hat \rho_0 $ and $ \hat \rho_3 $ which belong to even frequency symmetry class. These components belong to even-frequency spin-triplet odd-momentum even-band parity (ETOE) symmetry class. 
In Ref \cite{Asano:2015} the intraband pairing correlation is calculated as
\be
\label{Eq70}
F_{11, \uparrow  \downarrow } (k,i\omega _n ) - F_{11, \downarrow  \uparrow } (k,i\omega _n ) = \frac{{ - \Delta }}{{Z_5 }}i\omega _n V_3 ,
\ee
\be
\label{Eq71}
F_{22, \uparrow  \downarrow } (k,i\omega _n ) - F_{22, \downarrow  \uparrow } (k,i\omega _n ) = \frac{{\Delta }}{{Z_0 }}i\omega _n V_3 .
\ee
The hybridization generates pairing correlations that belong to the odd-frequency spin singlet odd-momentum even-band parity (OSOE) class.
As mentioned in Ref  \cite{Asano:2015} the interband pair correlation can be written as 
\be
\label{Eq72}
\begin{array}{l}
 [F_{12, \uparrow  \downarrow } (k,i\omega _n ) + F_{12, \downarrow  \uparrow } (k,i\omega _n )] + [F_{21, \uparrow  \downarrow } (k,i\omega _n ) + F_{21, \downarrow  \uparrow } (k,i\omega _n )] \\ 
  = \frac{{2\Delta }}{{Z_5 }}i\omega _n \xi _ -  . \\ 
 \end{array}
\ee
Thus the spin-orbit coupling generates pairing correlations that belong to the odd-frequency spin triplet even-momentum even-band parity (OTEE) class.
In contrast in our formalism the band asymmetry generates interband pairing correlation as 
\be
\label{Eq73}
[F_{12, \uparrow  \downarrow } (k,i\omega _n ) + F_{12, \downarrow  \uparrow } (k,i\omega _n )] + [F_{21, \uparrow  \downarrow } (k,i\omega _n ) + F_{21, \downarrow  \uparrow } (k,i\omega _n )] = 0 .
\ee
The interband pairing correlation due to band symmetry is 
\be
\label{Eq74}
\begin{array}{l}
 [F_{12, \uparrow  \downarrow } (k,i\omega _n ) + F_{12, \downarrow  \uparrow } (k,i\omega _n )] - [F_{21, \uparrow  \downarrow } (k,i\omega _n ) + F_{21, \downarrow  \uparrow } (k,i\omega _n )] \\ 
  = \frac{{ - 4\Delta }}{{Z_0 }}(\xi ^2  + \omega _n ^2  + \eta ^2 k^2 ) . \\ 
 \end{array}
\ee
These components belong to even-frequency spin-triplet even-momentum odd-band parity (ETEO) symmetry class.
Thus, for spin triplet, the spin dependent and spin independent hybridization both generate the same symmetry class ETEO due to interband pairing correlation. The odd frequency pairing arises in the presence of spin independent hybridization due to intraband pairing correlations. 

\section{Conclusion}
Within the theoretical model the existence of odd frequency pairs in two band superconductors by incorporating both spin independent hybridization and spin dependent spin-orbit interaction is investigated. This model also includes both the one-particle hybridization term and all possible intraband and interband superconducting pairing interaction terms in a two-band system.

The normal and anomalous thermal Green's functions have been calculated in the Nambu formalism as elements of the Fourier transformed $4 \times 4$ matrix Green's function by taking into account of all possible intraband and interband superconducting interaction terms coupling both bands in the mean field approximation.
By assuming that the attractive interaction acts on two electrons with different spins in different conduction bands different symmetry classes were demonstrated in the presence of hybridization and spin-orbit coupling.

The role of intraband and interband pairing correlations to emerge the odd frequency in a two-band superconductor was examined. For spin singlet, the odd-frequency is generated by spin dependent hybridization potential owing to intraband pairing correlations in agreement with the odd frequency generated by the interband pair correlation due to band asymmetry in Ref \cite{Asano:2015}. On the other hand, for spin triplet the spin independent hybridization potential generates the odd-frequency pairing due to intraband correlations in agreement with the result of Ref  \cite{Asano:2018}.


\providecommand{\href}[2]{#2}

\address{
Department of Physics, University of Isfahan,\\
Isfahan 81746, Iran\\
\email{h.yavary@sci.ui.ac.ir}\\
}

\end{document}